\documentclass[a4paper]{article}
\usepackage[utf8]{inputenc}
\usepackage[english]{babel}
\usepackage{amsmath}
\usepackage{graphicx} 
\usepackage{fancyhdr} 
\usepackage[colorinlistoftodos]{todonotes}
\usepackage{hyperref}
\usepackage{xcolor}
\usepackage{sectsty}
\usepackage{enumitem} 
\usepackage{csquotes}
\usepackage[T1]{fontenc}    
\usepackage{lettrine}       
\usepackage{yfonts}         
\usepackage{etoolbox}
\usepackage{url}
\usepackage{style}
\usepackage{booktabs}       
\usepackage{amsfonts}       
\usepackage{nicefrac}       
\usepackage{fontawesome}
\usepackage{caption}        
\usepackage[numbers,compress]{natbib}
\setlist[enumerate,1]{} 
\setlist[enumerate,2]{label*=\textcolor{boldgreen}\textbf{\alph*.}}  

\definecolor{mytitlecolor}{HTML}{267272}
\definecolor{boldgreen}{HTML}{3e9696} 
\definecolor{darkgray}{RGB}{130, 130, 130} 
\definecolor{lightgreen}{RGB}{144, 238, 144} 
\definecolor{lightred}{RGB}{235, 166, 160} 
\definecolor{blackbg}{RGB}{0, 0, 0} 

\newcommand{\greenbf}[1]{{\textcolor{boldgreen}{\bfseries#1}}}

\setlist[itemize]{label=\textcolor{boldgreen}{\textbullet}} 
\setlist[enumerate,1]{label=\textcolor{boldgreen}{\textbf{\arabic*.}}} 

\allsectionsfont{\color{mytitlecolor}}

\makeatletter
\patchcmd{\endabstract}{\vskip 1ex}{%
  \vskip 1ex%
  {\footnotetext{\faEnvelopeO\ \texttt{atboria@gmail.com}}}%
}{}{}
\makeatother

\title{Securing External Deeper-than-black-box\\ GPAI Evaluations}

\author{%
  Alejandro Tlaie \faEnvelopeO\\
  Independent \And Jimmy Farrell\\ Pour Demain
}

\begin{document}

\maketitle

\begin{abstract}

This paper examines the critical challenges and potential solutions for conducting secure and effective external evaluations of general-purpose AI (GPAI) models. With the exponential growth in size, capability, reach and accompanying risk of these models, ensuring accountability, safety, and public trust requires frameworks that go beyond traditional black-box methods. The discussion begins with an analysis of the need for deeper-than-black-box evaluations (\greenbf{Section \ref{part1}}), emphasizing the importance of understanding model internals to uncover latent risks and ensure compliance with ethical and regulatory standards. Building on this foundation, \greenbf{Section \ref{part2}} addresses the security considerations of remote evaluations, outlining the threat landscape, technical solutions, and safeguards necessary to protect both evaluators and proprietary model data. Finally, \greenbf{Section \ref{part3}} synthesizes these insights into actionable recommendations and future directions, aiming to establish a robust, scalable, and transparent framework for external assessments in GPAI governance.

\end{abstract}

\newpage
\tableofcontents
\newpage

\newpage

\section*{Executive Summary}\label{exec_sum}
\addcontentsline{toc}{section}{Executive Summary}

The rapidly developing capability and pervasiveness of frontier general purpose AI (GPAI) models present critical challenges for accountability, safety, and public trust. Third-party external assessments are an indispensable component of any robust risk assessment and mitigation framework. They help identify a broader range of risks and maintain public confidence in AI technologies by providing independence, methodological diversity, and credibility—capabilities that internal evaluations alone often cannot deliver.

The current paradigm of external black-box GPAI evaluation techniques mostly relies on effectively asking a machine whether it is safe, rather than rigorously verifying its safety through more objective means \cite{bengio2025international, mallen2023eliciting, burns2022discovering, bucknall2023structured}. This can be roughly compared to conducting biological experimentation without a microscope; only observing results with the naked eye. Whilst numerous recent examples of black-box independent external assessments have shown promising results in identifying risks \cite{METR-1,meinke2024frontier} this constrained evaluation method struggles to address certain complex model propensities \cite{hubinger2024sleeper,panickssery2023reducing,Farquhar2024Detecting,azaria2023internal}. A deeper-than-black-box approach, with appropriately expanded access for external assessors, is therefore essential to address systemic risks effectively and uphold rigorous safety protocols.

\subsection*{Key Challenges}
\begin{enumerate}
    \item \greenbf{Limited Access for External Evaluations:} Although the science of GPAI evaluation is advancing quickly—from black-box (input/output only) to full white-box (complete access to model internals)—current external assessments are most often confined to black-box scenarios. This restricted view can obscure critical insights into how models generate outputs and where they might harbor vulnerabilities. Different levels of access are required to evaluate different safety dimensions, yet existing practices and policies have not kept pace with the evolving risk landscape.
   \item  \greenbf{Security and Confidentiality Risks:} Remote evaluations are vulnerable to cyber threats, including model theft, manipulation, and data exfiltration. Nation-states and other advanced actors pose significant risks in an unregulated landscape.
   \item \greenbf{Global Governance and Standards Gaps:} While certain jurisdictions, such as the EU, are pioneering regulatory initiatives (e.g., the EU AI Act) that can influence global market practices (“the Brussels Effect”), there remains a lack of universally recognized frameworks or standards for conducting secure and thorough third-party AI assessments. Unlike in industries such as nuclear energy—where recognized international guidelines and national regulatory bodies coordinate on safety standards—AI is still in the early stages of establishing broadly accepted benchmarks for safe external evaluations.
\end{enumerate}

\subsection*{Solutions and Recommendations}
\begin{enumerate}
    \item \greenbf{Deeper-than-Black-Box Assessments:} To address systemic and complex risks, external evaluations must go beyond black-box approaches. Controlled but substantial access to model internals, training data lineage, and other technical details is crucial for understanding AI behavior, identifying hidden vulnerabilities, and mitigating risks under adversarial conditions.
    \item \greenbf{Cyber Threat Identification \& Safeguards:} A layered defense strategy is needed to protect against insider threats, cyberattacks, and process vulnerabilities. Threat modeling should classify risks according to organizational capacity and intentionality, ensuring that safeguards align with the sophistication of potential attackers. Technological solutions (e.g., secure enclaves, differential privacy), physical safeguards (e.g., on-site evaluations under strict protocols), and legal measures (e.g., NDAs, contractual arrangements) form an integrated toolkit to uphold data confidentiality and process integrity. 
    \item \greenbf{Standards, National Oversight \& Global Collaboration:} While an international regulatory body for AI may not be realistic in the near term, rapid progress can be made through:
    \begin{enumerate}
        \item \underline{Preliminary Guidelines:} Develop and share initial best practices for external AI evaluation, focusing on consistency, security, and auditability.
        \item \underline{Global and Cross-Industry Standards:} Encourage standards bodies, professional associations, and multilateral forums (e.g., ISO, IEEE) to codify technical protocols and methodologies for AI audits and model evaluations.
        \item \underline{National Oversight Functions:} Establish or strengthen national authorities (e.g., specialized AI safety institutes) to certify or accredit third-party evaluators, ensuring rigorous, context-specific oversight that aligns with emerging global guidelines.
    \end{enumerate}
Beyond the immediate steps of creating guidelines and standards, longer-term multilateral coordination remains vital. A decentralized but cooperative platform could help standardize access protocols, model-evaluation pipelines, and data-sharing safeguards. Such infrastructure, complemented by bilateral or multilateral agreements, can help manage geopolitical concerns and unify safety efforts, even in the absence of a single global regulatory body. 
\end{enumerate}

\newpage

\section{Opening the black-box}\label{part1}

\begin{quote}
``\textit{We can only see a short distance ahead, but we can see plenty there that needs to be done.}''

-Alan Turing, 1950.
\end{quote}

\subsection{The need for external evaluations with deeper model access}

\lettrine%
  {\raisebox{-0.2\baselineskip}{\includegraphics[height=1.5em]{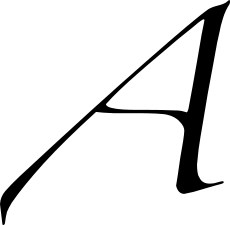}}}%
  {s} frontier GPAI models proliferate furtherinto society, the question of \greenbf{accountability and trust} becomes increasingly urgent. Beyond simply identifying areas such as bias, misuse potential, or misalignment, \greenbf{external evaluations} stand apart by mitigating conflicts of interest inherent in internal reviews. While internal assessments may offer proprietary insights, they are often constrained by \greenbf{organizational biases} and commercial priorities. By contrast, \greenbf{independent external reviewers} add methodological diversity and credibility, uncovering blind spots and enhancing public trust. This shift away from the era of \greenbf{self-regulation}—where market incentives conflict with societal interests—marks a pivotal evolution in governance, as recognized by frameworks like the Digital Services Act ((C(2023) 6807, Article 3)) and the EU AI Act’s Code of Practice \cite{EU2024AI}.

There is ample precedent for this kind of oversight in other high-stakes sectors. For instance, the Digital Services Act mandates rigorous audits of Very Large Online Platforms, including comprehensive access to source code and all data relevant for the audit. Similarly, the EU Medical Device Regulation requires external conformity assessments for high-risk devices, often involving reviews of algorithms, training data, and validation protocols. These examples demonstrate that \greenbf{external scrutiny} is both feasible and vital to mitigate risks for technologies with significant societal impact.

A common critique of mandatory external assessments is that the necessary infrastructure or ecosystem does not yet exist—a valid concern that nevertheless underscores the \greenbf{urgency of moving forward}. Early decisions about auditing standards will shape industry norms. Delaying robust external evaluation requirements risks entrenching suboptimal or opaque practices that will be far more difficult to reform later. A tiered approach—focusing first on high-stakes applications or frontier models—can ensure feasibility without compromising the core goal of \greenbf{strengthening AI accountability}.

This important shift away from the era of self-regulation by AI developers, where market incentives conflict heavily with societal needs, is not unprecedented. Lessons from fields like cybersecurity \cite{egloff2021attribution} and finance \cite{elshandidy2021independent} demonstrate the importance of \greenbf{independent oversight}, especially for high-stakes technologies. However, a significant challenge in AI governance is the profound uncertainty surrounding a deep understanding of how these models actually work. This \greenbf{opacity} further reinforces the necessity of external assessments that go beyond black-box evaluations to uncover deeper structural risks and capabilities.

Frontier GPAI models introduce unique challenges from an evaluation perspective. A major issue is that a model’s behavior may vary depending on its deployment context—a limitation that is particularly problematic for black-box evaluations, where external testing cannot capture the full spectrum of possible outputs. However, if provided with \greenbf{deeper-than-black-box (or white-box) access}, evaluators can directly inspect the model’s inner workings (neuronal activations, attention heads, residual stream,...), revealing latent harmful capabilities independent of deployment. This approach: \greenbf{I)} mitigates the risks associated with context-dependent behavior and \greenbf{II)} addresses the choke-point issue inherent in the value chain, where providers hold significant control over deployment parameters. Given that even developers have limited insight into the model’s internal mechanisms, these deeper external assessments are crucial for identifying and mitigating unforeseen vulnerabilities post-deployment.

Nevertheless, significant challenges remain. One major issue is the difficulty in assessing risks independently of the deployment context. The same GPAI model can display vastly different behaviors depending on its application, complicating any absolute “inherent risk” assessment. Additionally, the \greenbf{opacity} of these models—even to their developers—amplifies the threat of unforeseen capabilities or vulnerabilities emerging post-deployment. \greenbf{Deeper-than-black-box access} can help external evaluators conduct more exhaustive testing, detect \greenbf{latent weaknesses} (such as hidden backdoors or potential for fine-tuning), and accelerate independent safety research. In turn, this reduces the likelihood of missed risks and addresses \greenbf{accountability gaps} when harm or misuse occurs—a particularly thorny issue given the often unclear division of responsibility between technology providers and downstream developers.

Compounding these concerns, \greenbf{frontier GPAI development}, including the beginning of the inference-time compute paradigm \cite{jaech2024openai} and release of open-weights reasoning models \cite{guo2025deepseek}, \greenbf{is increasingly outpacing governance efforts} and technical oversight methodologies. Traditional application-level checks are insufficient to address systemic issues, such as undisclosed biases or vulnerabilities that might only manifest under rare or adversarial conditions. To bridge the gap between governance aspirations and technical realities, new forms of \greenbf{supervision, oversight, and control} are necessary throughout the AI lifecycle. These must include robust methodologies for understanding model internals, detecting latent vulnerabilities, and verifying compliance with ethical and legal standards from the training phase onward. Application-level evaluations must be an addition to \greenbf{deeper, context-sensitive, and methodologically diverse external assessments}, so that regulators and stakeholders can ensure that GPAI models evolve under the twin imperatives of safety and accountability.

\subsection{SOTA Evaluations Based on Different Levels of Access}

In the context of foundation models, state-of-the-art (SOTA) external evaluation methods must be adapted to accommodate varying levels of access \cite{casper2024black}. These levels include:
\begin{itemize}
    \item \greenbf{Black-box:} Access to inputs, model queries, and outputs.
    \item \greenbf{Grey-box:} Limited access to model internals, such as input embeddings, hidden neural activations, or sampling probabilities.
    \item \greenbf{De facto white-box:} Ability to indirectly run arbitrary processes on the system without copying parameters.
    \item \greenbf{White-box:} Full access.
\end{itemize}

In this section, we outline state-of-the-art (SOTA) evaluation techniques based on varying levels of access to the model. Each technique will be detailed with the following structure: first, the objective of the method will be introduced, explaining its purpose within the evaluation framework. Next, we will describe the specific methodologies employed, highlighting key tools and approaches. Finally, the value of each technique to external evaluations will be clarified, emphasizing its role in uncovering risks, ensuring compliance, and enhancing trust in AI systems. By structuring the discussion in this way, we aim to provide a clear and comprehensive understanding of how these techniques can address the unique challenges posed by foundation models.

Thus, depending on which type of access is provided, evaluators interact with the AI model using different means; among others: via sandboxed environments or encrypted APIs or using advanced interpretability techniques that allow for a deeper understanding of the model’s inner workings.

\begin{figure}[h]
\centering
\includegraphics[width=\textwidth]{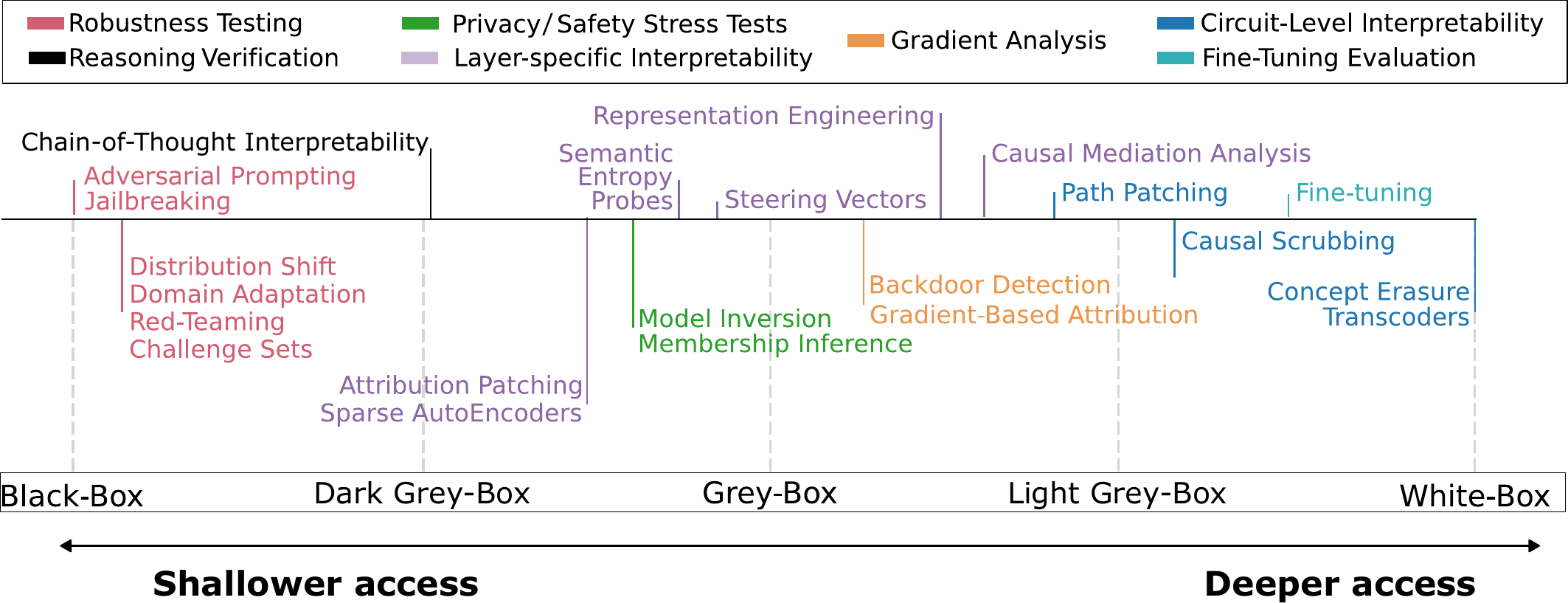}
\caption{\greenbf{Different access-specific AI evaluation techniques.} From completely black-box access (left-most part of the diagram) to completely white-box access (right-most part), there is a wide variety of SOTA techniques that an external evaluator could perform depending on the level of model access that they are granted. We colour-code techniques belonging to the same group (see in-figure legend), even if there is a clear ordering of model access that is needed to perform each of these (e.g. Robustness Testing (red) requires almost completely black-box access, while Circuit-Level Interpretability (dark blue) needs white-box access).
}
\label{fig:access-evaluation}
\end{figure}

\subsubsection{Mechanistic Interpretability}

\underline{Objective:} Understand and verify how the model internally represents concepts, reasons through inputs, and arrives at outputs.
\paragraph{Layer-specific interpretability (dark-to-light grey-box):} Even if only layer-specific information was provided to the evaluator, there are several techniques that would allow them to comprehensively inspect the model internals, on a layer-by-layer basis:
\begin{itemize}
    \item \greenbf{Sparse AutoEncoders:} Likely one of the most promising interpretability techniques. They learn sets of sparsely activating features that are more interpretable and monosemantic than directions identified by alternative approaches, where interpretability is measured by automated methods. It’s been shown that we can pinpoint the features that are causally responsible for counterfactual behaviour in certain tasks \cite{Cunningham2023Sparse}. Sparse AutoEncoders were notably scaled up to a production model (Claude 3 Sonnet) \cite{templeton2024scaling}.
    \item \greenbf{Representation Engineering:} Rather than modifying the prompt or weights of the LLM, the model’s behaviour can be directly controlled by intervening on its activations during a forward pass \cite{wehner2023representation}. This allows for highly targeted testing of how specific representations influence behaviour.
    \item \greenbf{Semantic Entropy Probes:} Semantic entropy probes have been developed and tested \cite{Farquhar2024Detecting} to measure the hallucination rate of several LLMs. These probes evaluate the internal activations of models to determine whether they represent coherent, human-readable semantic structures across layers. Evaluators could use these to quantify alignment between expected and actual representations in tasks like summarisation or reasoning.
    \item \greenbf{Causal Mediation Analysis:} Recently applied \cite{Vig2020Causal} to detect gender bias in smaller language models, this technique identifies causal pathways within models by measuring how specific interventions on activations mediate changes in outputs. It can be extended to investigate biases or the influence of undesirable latent concepts.
    \item \greenbf{Steering Vectors:} Linear directions (vectors) can be introduced in the model’s residual stream \cite{Burns2023Steering, arditi2024refusal}. These vectors shift the model’s behaviour in predictable ways, such as promoting cautiousness or fairness. Evaluators can apply these vectors and quantify the resulting changes to output distributions, providing insight into the model’s responsiveness to ethical interventions.
    \item \greenbf{Attribution Patching:} A lower-cost variant of activation patching (see below), attribution patching \cite{Nanda2023Attribution} is ideal for iterative debugging during evaluations. It allows evaluators to test whether specific attribution changes alter outputs without needing full model access.
\end{itemize}

\paragraph{Circuit-level interpretability (dark-to-light grey-box):} With more comprehensive access, neural networks can be decomposed into interpretable “circuits” of neurons or attention heads. By mapping these functional subunits, evaluators can identify whether the model encodes meaningful and coherent representations of concepts and logic. 

\begin{itemize}
    \item \greenbf{Path Patching:} Authors have localised particular behaviours to specific paths within models \cite{goldowsky2023localizing}. This allows evaluators to isolate the causal paths for specific outputs, improving their ability to assess responsibility and alignment for behaviours.
    \item \greenbf{Causal Scrubbing:} Used to test interpretability hypotheses, causal scrubbing involves behaviour-preserving resampling ablations \cite{Elhage2022Causal}. It rigorously tests whether specific hypotheses about model internals align with observed behaviours.
    \item \greenbf{Transcoders:} These tools decompose model computations into interpretable circuits \cite{dunefsky2024transcoders}, enabling evaluators to verify the modularity of computation and the semantic roles of individual subunits. 
    \item \greenbf{Concept Erasure:} This technique removes specified features from a representation \cite{Ravfogel2023LEACE}. It has dual utility: improving fairness (e.g., preventing a classifier from using protected attributes like gender or race) and enabling evaluators to test the causal role of a concept by erasing it and observing changes in behaviour. 
\end{itemize}

\subsubsection{Robustness Testing}

\underline{Objective:} Evaluate the model’s resilience to adversarial perturbations, domain shifts, and malicious attempts to exploit vulnerabilities.

\begin{itemize}
    \item \greenbf{Adversarial Example Generation (dark-grey-box):} Systematically generate inputs 
    \cite{Rauber2017Foolbox,croce2020reliable} designed to confuse or mislead the model. If the model fails under slight perturbations, it indicates brittle reasoning. \underline{Value to external evaluations:} A neutral evaluator can assess how easily the model can be tricked into producing harmful or nonsensical outputs. High vulnerability suggests that the model might be unsafe for deployment, even if the evaluator never sees the original training data.
    \item \greenbf{Distribution Shift/Domain Adaptation Tests (black-box):} Evaluate performance on data from different domains or with altered characteristics (e.g. through the WILDS Benchmark \cite{koh2021wilds}). \underline{Value to external evaluations:} The evaluator can verify if the model’s claimed robustness holds in new contexts. This assures stakeholders that the model won’t fail catastrophically when facing diverse user populations or novel conditions.
    \item \greenbf{Red-Teaming (black-box):} Present the model with curated inputs designed to test its guardrails or identify vulnerabilities in compliance mechanisms (e.g., requests for disallowed content). This is already common practice in the industry \cite{OpenAI2023RedTeaming,ganguli2022red}. \underline{Value to external evaluations:} Through controlled, reproducible stress tests, an external evaluator evaluates whether the model maintains its built-in safeguards and resists attempts to bypass or remove them. This can reassure regulators and the public that the model cannot be easily manipulated to produce harmful or unintended outputs.
\end{itemize}

\subsubsection{Gradient Analysis}

\underline{Objective:} Investigate how gradients (sensitivity of outputs to changes in inputs or parameters) reveal hidden biases, training irregularities, or backdoors.

\begin{itemize}
    \item \greenbf{Backdoor Detection via Gradient Patterns (grey to light-grey-box):} Identify if certain input triggers (e.g., specific phrases or patterns) cause the model to behave unexpectedly, indicating a hidden backdoor \cite{zhu2023gradient,pan2023asset}. \underline{Value to external evaluations:} Even with no direct access to training code, an external evaluator can discover stealthy tampering or malicious implants that compromise model trustworthiness.
    \item \greenbf{Gradient-Based Attribution (grey to light-grey-box):} Determine which features are most influential in the model’s decision \cite{sundararajan2017axiomatic,smilkov2017smoothgrad}. If sensitive attributes consistently rank high, it suggests discriminatory behaviour. \underline{Value to external evaluations:} Without proprietary data, an evaluator can still quantify and highlight implicit bias, providing objective evidence of fairness issues that require remediation.
\end{itemize}

\subsubsection{Fine-Tuning Evaluation}

\underline{Objective:} Assess the safety and integrity of a model after it has been adapted or specialized for new tasks or domains.

\begin{itemize}
    \item \greenbf{Before-and-After Safety Comparisons (light-grey-box):} Compare model metrics on fairness, robustness, and compliance pre- and post-fine-tuning. \underline{Value to external evaluations:} An evaluator can confirm that modifications do not necessarily degrade safety or introduce vulnerabilities. This is critical when external updates (like domain adaptation) might have unintended negative effects.
\end{itemize}

\subsubsection{Privacy \& Safety Stress Tests:}

\underline{Objective:} To ensure that the model does not leak training data, personal information, or proprietary content.

\begin{itemize}
    \item \greenbf{Membership Inference and Model Inversion Attacks (grey-box):} Tests whether information about specific data points in the training set or reconstructed training samples can be inferred from the model's outputs \cite{hu2022membership,zhou2024model}. \underline{Value to external evaluations:} These tests assess the risk of data leakage from the model, providing evidence of whether the system meets data confidentiality requirements. Evaluators can evaluate compliance with privacy laws and standards without needing access to the underlying training data.
\end{itemize}
 
\subsubsection{Reasoning Verification}

\underline{Objective:} Validate that the model’s reasoning steps are sound, coherent, and aligned with intended logic.

\begin{itemize}
    \item \greenbf{Chain-of-Thought evaluations (dark-grey-box):} Examine the model’s step-by-step reasoning (Chain-of-Thought, CoT). Recent efforts have started using Mechanistic Interpretability techniques to CoTs \cite{dutta2024think}, which may become more relevant in the near-term as the paradigm shifts from relying on pre-training to test-time compute. \underline{Value to external evaluations:} Ensures the evaluator can verify rationality and detect hallucinations, even when they cannot inspect the raw training code.
\end{itemize}

\section{Guaranteeing security}\label{part2}

\subsection{On-site evaluations: the preferred but currently unfeasible format}\label{onsite}

When it comes to secure evaluations, on-site assessment are the preferred option because they enable full control over access to sensitive AI systems, including proprietary model weights and training data. They reduce the risk of data exfiltration and ensure compliance with jurisdiction-specific requirements, providing direct oversight for thorough evaluations.

However, legal and geopolitical complexities currently make on-site evaluations challenging without a global governance framework. Even if there are analogous challenges to the ones faced in the nuclear domain, there is no global authority for AI oversight akin to the International Atomic Energy Agency (IAEA). Nation-states may view on-site evaluations as intrusions into their sovereignty and fear espionage, leading to significant resistance to allowing foreign entities access to sensitive AI infrastructure.

Consequently, we advise national governments and supra-national organisations to reach agreements constituting such a regulatory body. In the meantime, there are already technically feasible alternatives like voluntary on-site evaluations by foreign entities, akin to the current paradigm–but with deeper-than-black-box access. In the interim period before such international governance frameworks introduce mandatory obligations, voluntary on-site third party evaluations may lay the groundwork for functional, cooperative and mutually beneficial dialogue between GPAI model providers and regulators.

\subsection{Threat landscape of a remote evaluation}

Remote evaluations offer a scalable alternative to on-site inspections for AI systems but face significant security challenges, such as safeguarding sensitive model components and mitigating risks of theft or misuse. This section categorizes threats to remote AI evaluations and maps them to corresponding \greenbf{Operational Capacity (OC)} levels, closely following \cite{nevo2024securing}. These levels range from \greenbf{OC1} (hobby hackers and untargeted attacks) to \greenbf{OC5} (highly resourced state-sponsored campaigns), with room to accommodate intermediate threats: OC2 (professional opportunistic hackers), OC3 (cybercrime orgs/insider threats) and OC4 (standard ops by leading cyber-capable institutions). Furthermore, we added a second axis (\greenbf{Intentionality}, also ranging from \greenbf{I1} to \greenbf{I5}), that accounts for the degree of the need for a motivated actor for each threat to materialize. We believe that this mapping is useful to tailor mitigation strategies that are specific to each (OC, I) level. Next to each of the examples we have identified as attack vectors (that are both general cyberthreats and AI-specific), we outline concrete scenarios to illustrate how these threats could manifest.

\begin{figure}[h]
\centering
\includegraphics[width=0.8\textwidth]{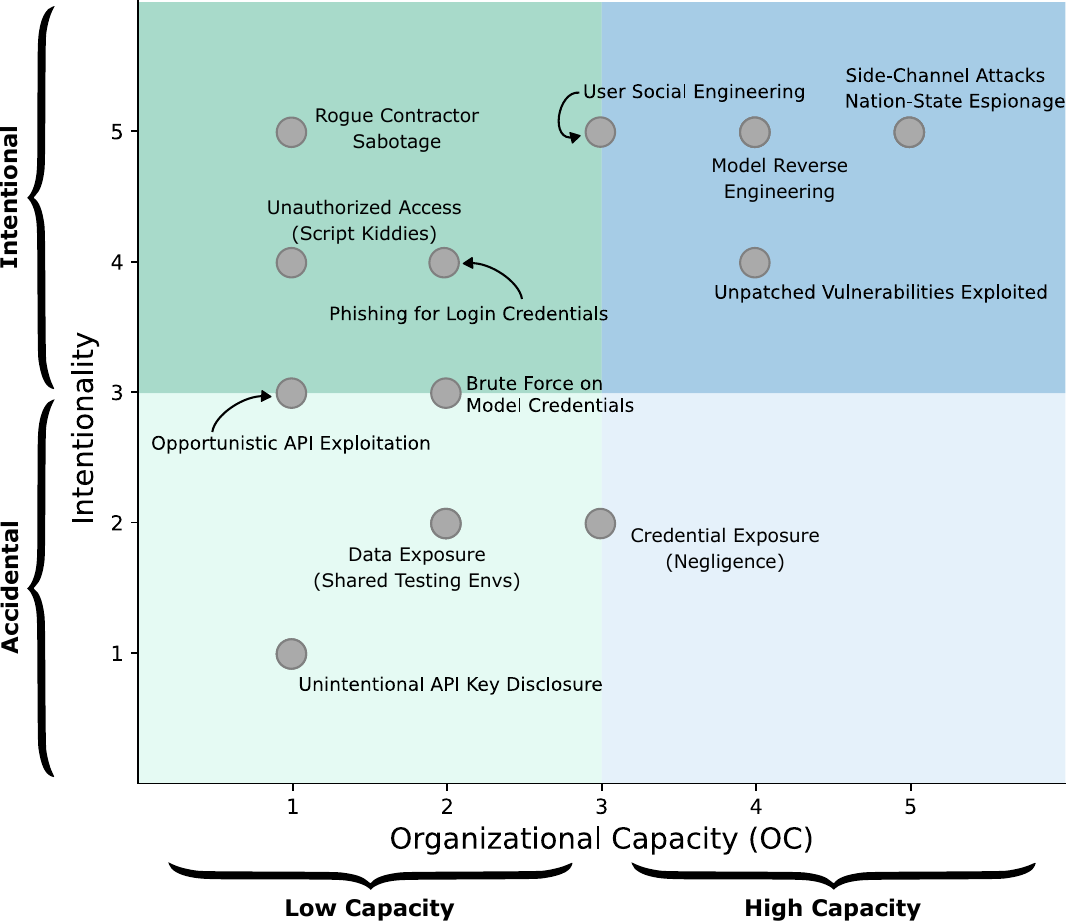}
\caption{\greenbf{example scenarios from the threat landscape when performing remote AI model evaluations.} We map different relevant threats onto two axes: Organizational Capacity (following work from \cite{nevo2024securing}) and Intentionality (to which degree the relevant attack vector needs a motivated actor to take place).
}
\label{fig:threat-landscape}
\end{figure}

\begin{enumerate}
    \item \greenbf{\underline{External Threats:}} Adversaries external to the organization target remote evaluation systems to exfiltrate proprietary data, manipulate results, or disrupt the process. These threats, can be illustrated through the following scenarios:
    \begin{enumerate}
        \item \greenbf{Opportunistic API Exploitation (OC1, I3):} An adversary systematically queries the remote evaluation API to extract model behaviour or reverse-engineer sensitive components. This tactic is particularly effective when evaluation systems expose fine-grained query responses without sufficient rate-limiting or output obfuscation.
        \item \greenbf{Unauthorized Access by Script Kiddies (OC1, I4):} Although limited in skill, script kiddies deliberately seek to breach remote evaluation systems using readily available exploit scripts or tools.
        \item \greenbf{Phishing for Login Credentials (OC2, I4):} Moderately advanced attackers craft targeted phishing campaigns to trick evaluators into revealing remote evaluation credentials, enabling unauthorized model access or data exfiltration.
        \item \greenbf{Nation-State Espionage (OC5, I5):} A highly resourced government-sponsored program infiltrates the evaluating infrastructure—potentially via zero-day exploits—to steal proprietary model weights or manipulate evaluation processes.
        \item \greenbf{Model Reverse Engineering (OC4, I5):} Advanced threat actors intercept or query the model in ways that reveal its architecture or training data. With significant resources, they fine-tune attack strategies to reverse-engineer valuable intellectual property.
    \end{enumerate}
    \item \greenbf{\underline{Insider Threats:}} Individuals with privileged or direct access to the evaluating process can misuse their roles—whether deliberately or negligently—for unauthorized activities:
    \begin{enumerate}
        \item \greenbf{Credential Exposure by Negligence (OC3, I2):}  An evaluator carelessly stores remote evaluation credentials in an unsecured repository, inadvertently granting external parties access to sensitive systems.
        \item \greenbf{Rogue Contractor Sabotage (OC1, I5):} A disgruntled or malicious contractor, despite limited organizational capacity, is highly motivated to disrupt or leak confidential information from within the evaluating process.
    \end{enumerate}
    \item \greenbf{\underline{Process and Infrastructure Vulnerabilities:}} Distributed remote evaluating environments often involve multiple geographic locations and vendors, exposing various infrastructure weaknesses:
    \begin{enumerate}
        \item \greenbf{Unintentional API Key Disclosure (OC1, I1):} An API key is accidentally posted or logged in a public-facing environment, allowing anyone who discovers it to query or tamper with remote evaluation services.
        \item \greenbf{Data Exposure via Shared Testing Environments (OC2, I2):} When testing resources are shared across different projects or teams, a misconfiguration can inadvertently leak evaluation data to unauthorized users.
        \item \greenbf{User Social Engineering Exploits (OC2, I3):} Attackers with moderate skills manipulate end-users or evaluators (through spear-phishing or social engineering) to gain unauthorized access to remote evaluation systems.
        \item \greenbf{Unpatched Vulnerabilities Exploited (OC4, I4):} Threat actors with significant resources exploit known but unpatched software or hardware vulnerabilities in remote evaluating platforms to access or manipulate the evaluating process.
    \end{enumerate}
    \item \greenbf{\underline{Inference and Leakage Risks:}} Risks arise when attackers exploit the outputs or logs of the evaluation process to glean sensitive details:
    \begin{enumerate}
        \item \greenbf{Side-Channel Attacks (OC5, I5):}   Elite adversaries use hardware or timing side-channels to collect subtle signals (e.g., power consumption, timing information), thereby inferring proprietary aspects of the model or its training data \cite{dutta2023spy,Naghibijouybari21side}.
    \end{enumerate}
\end{enumerate}

We believe that addressing these threats requires a layered approach that combines advanced technical safeguards with robust governance frameworks. However, in this report we will focus mostly on the technical mitigations that either already exist or are expected to become state-of-the-art in the very near term. In Section \ref{legal}, we outline some Legal Safeguards that should be considered jointly with the technical mitigations we propose.

We follow the work of others \cite{trask2020beyond} in framing GPAI external evaluations as an information flow problem: how to guarantee that only the required information goes from the sender to only the authorized receiver? As previous authors have argued \cite{tramer2022position}, it is crucial that technical solutions focus on both preventing unauthorized access and limiting what can be learned from the evaluation process itself.

\begin{figure}[h]
\centering
\includegraphics[width=1\textwidth]{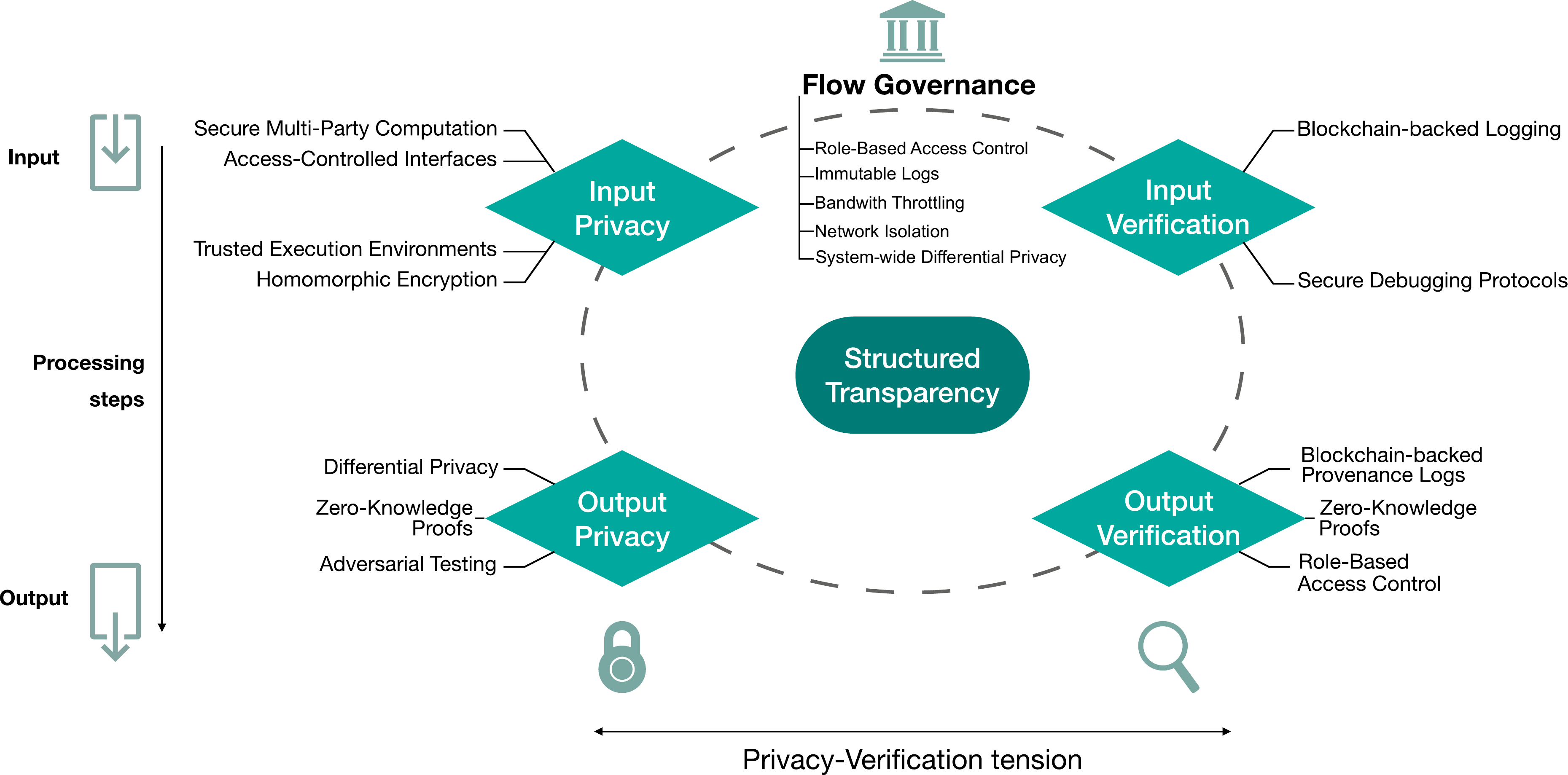}
\caption{\greenbf{a graphical summary of how the Structured Transparency framework \cite{trask2020beyond} can look in practice.} There are five key elements to guaranteeing that the flow of information goes only from the sender to the target receiver. For each aspect of the framework, we list all of the mitigations we explain in the main text.
}
\label{fig:struct_trans}
\end{figure}

\subsection{Technical Solutions}\label{cybersec}

Emerging initiatives like OpenMined\footnote{\url{https://openmined.org}}, the National Deep Inference Fabric (NDIF)\footnote{\url{https://ndif.us}}, or NNsight\footnote{\url{https://nnsight.net/}} exemplify the advancement of technical solutions aimed at enhancing privacy and transparency in AI systems. These projects provide frameworks and tools that enable secure data analysis and facilitate in-depth examinations of large-scale AI models without compromising sensitive information. Their existence underscores the availability of technical solutions that can enhance privacy and transparency in external audits of General-Purpose AI (GPAI) systems, moving beyond traditional black-box evaluations.

\subsubsection{Input Privacy}

\underline{Core concern:} When evaluating remotely, data owners must share information or allow computations without losing control. These tools ensure that sensitive inputs (such as proprietary model parameters or private user data) are never exposed, even when processed off-site.

\begin{itemize}
    \item \greenbf{Homomorphic Encryption (HE):} In a remote evaluation, model owners and evaluators may operate in different organizations or jurisdictions. Normally, sending unencrypted data to the evaluator risks leaks or unauthorized use. With HE, computations (e.g., fairness checks, accuracy tests) occur on encrypted data. This means the evaluator never sees the unencrypted model parameters or training samples. The remote nature of the evaluation, which often involves sending data across the internet or working in shared cloud environments, makes HE’s confidentiality especially important. It ensures sensitive proprietary information or personal data cannot be intercepted or misappropriated at any point in the remote pipeline. Nevertheless, a necessary caveat is that this technology is far from being ready to use in modern LLMs \cite{tramer2022position}.
    \item \greenbf{Secure Multi-Party Computation (SMPC):} In many remote evaluations, multiple stakeholders contribute data and expect evaluation results. SMPC allows each party to keep their inputs secret while still collectively performing meaningful calculations \cite{gamiz2025challenges}. For instance, multiple companies might jointly assess their combined training sets for bias or safety concerns without revealing their individual datasets. In a remote context where trust is low and geographical separation is high, SMPC provides a way to collaborate securely across distances. It mitigates risks of collusion or accidental data leaks by ensuring no single remote participant sees the other’s raw inputs.
    \item \greenbf{Trusted Execution Environments (TEEs):} A TEE creates a “black box” on a remote machine where code and data can be processed securely, shielded from the host system’s operators. In a remote evaluation, you don’t physically control the hardware your data runs on. TEEs ensure that even if the remote host is compromised or curious, the computations remain inaccessible and tamper-resistant. This hardware-backed guarantee \cite{islam2024confidential} helps bridge the trust gap that distance and external infrastructure can create. Based on the current evidence \cite{tramer2018slalom}, we believe this is the most promising technical solution to secure input privacy; however, it is currently unknown whether TEEs can handle billion-scale parameter models.
    \item \greenbf{Access-Controlled Interfaces:} When conducting remote evaluations, data requests are typically made through APIs or secure web portals. By strictly limiting the types of queries allowed and the data accessible through these interfaces, organizations can significantly reduce the risk of evaluators or unauthorized users “fishing” for sensitive information. This approach is particularly relevant in light of recent research \cite{carlini2024stealing}, we believe this is the most promising technical solution to secure input privacy; however, it is currently unknown whether TEEs can handle billion-scale parameter models.
\end{itemize}

\subsubsection{Output Privacy}
\underline{Core concern:} Even if the evaluator only receives summarized results, there is a risk that the outputs can be reverse-engineered to reveal sensitive data or model details. Output privacy techniques ensure that what is learned from the evaluation results cannot compromise confidentiality.

\begin{itemize}
    \item \greenbf{Differential Privacy:} Remote evaluations often produce performance metrics or bias reports. Without safeguards, these outputs can be analyzed to infer sensitive training examples or model structure. Differential privacy \cite{dwork2014algorithmic} adds statistical noise to mask individual-level details. This protects against an evaluator who, operating remotely and possibly out of direct oversight, tries to glean hidden patterns from the aggregate results. Even if network eavesdroppers intercept the results in transit, differential privacy ensures they cannot extract sensitive insights.
    \item \greenbf{Zero-Knowledge Proofs (ZKPs):} ZKPs enable proof of compliance (e.g., that a model meets a certain criterion) without revealing the underlying data or code. In a remote scenario, model owners want to show regulators or evaluators that certain standards are met, but they fear revealing their intellectual property. ZKPs confirm the truth of statements about the model without sharing its internals \cite{Ronis2024ZeroKnowledge}. This is invaluable in a globalized environment where evaluators and evaluated parties may be strangers to each other, yet still need a mechanism to trust and verify without disclosure. Some concrete proposals to evaluate Machine Learning models have already been put forward \cite{waiwitlikhit2024trustless}, but it remains to be seen how these scale up to state-of-the-art LLMs.

    \item \greenbf{Adversarial Testing in Sandboxed Environments:} When evaluators run tests remotely, the environment in which the model is tested is critical. By placing all testing in a secure sandbox—an isolated environment with strict controls—one ensures that even when pushing the model with tricky or adversarial inputs, no sensitive data leaks out. This protects against malicious testers who might attempt to coerce the model into revealing proprietary details through crafted queries. The sandbox containment ensures that no matter where in the world the evaluator is, they’re operating in a controlled, locked-down digital space.
\end{itemize}

\subsubsection{Input Verification}

\underline{Core Concern:} Remote evaluating often involves sending models and data to third parties or hosting them on external platforms. Without physical access and face-to-face verification, ensuring the authenticity and integrity of the inputs used in the evaluation is crucial.

\begin{itemize}
    \item \greenbf{Blockchain-Backed Logging:} By recording the submission of data and models on a blockchain, an immutable, tamper-evident timeline is created. In a remote evaluation scenario, it’s challenging to confirm that the model tested is the same one initially provided. Blockchain ensures no party can quietly swap out models or alter data retrospectively. Everyone can remotely verify that what gets evaluated is exactly what was intended, and discrepancies are easily detectable. It is important to note that this is a technically feasible approach and that it is supported by current research \cite{shi2022auditem,feng2025blockchain}. However, practical implementation requires addressing challenges related to scalability and integration to ensure the system is effective and secure.
    \item \greenbf{Secure Model Debugging Protocols:} Before running the evaluation, these protocols check digital signatures or cryptographic hashes of the submitted model. Even at a distance, one can confirm that the model has not been tampered with. This prevents scenarios where a remote party tries to substitute a less complex or sanitized model into the evaluating pipeline, undermining the evaluation’s credibility.
\end{itemize}

\subsubsection{Output Verification}

\underline{Core Concern:} Just as inputs need trust, the reported results also need validation. In remote evaluations, where results may be published to stakeholders scattered across different locations, ensuring that the final reports are accurate and not manipulated en route is essential.

\begin{itemize}
    \item \greenbf{Zero-Knowledge Proofs (ZKPs):} ZKPs don’t just protect input data \cite{Ronis2024ZeroKnowledge}—they can also validate that the reported conclusions (like “the model passed a certain fairness threshold”) are correct without revealing sensitive calculation details. This assures remote stakeholders that the published findings are not falsified. They can trust the outcome even if they never see the raw data or computations.
    \item \greenbf{Blockchain-Backed Provenance Logs:} Just as blockchain secures inputs, it can also record the entire evaluating process and outputs. In a remote context, results may be transmitted over various networks and time zones. By anchoring them to a blockchain, anyone can verify that the reported metrics and conclusions match what was generated during the evaluation. Any attempt at altering results afterwards would be evident, building confidence in remote processes.
    \item \greenbf{Role-Based Access Control (RBAC):} In a remote team, not everyone should have the authority to finalize or distribute the evaluation results. RBAC ensures that only authorized individuals can produce or publish verified conclusions. This reduces the risk of internal sabotage or error when the collaborators never meet in person, as trust must come from strict and enforceable digital policies.
\end{itemize}

\subsubsection{Flow Governance}

\underline{Core Concern:} Beyond just inputs and outputs, governing the entire evaluating process is challenging when people, servers, and data centres might be spread across continents. Flow governance tools ensure that every information transfer, action, and computation obeys security and compliance standards.

\begin{itemize}
    \item \greenbf{Role-Based Access Control (RBAC) in Governance:} Assigning roles and permissions is critical when participants are working remotely. The evaluator might be external, the model owner might be elsewhere, and the infrastructure could be managed by a third party. RBAC ensures that each party’s access is strictly defined. This prevents remote actors from overstepping their boundaries or accessing areas they should not, fostering a structured and safe environment.
    
    \item \greenbf{Network Isolation and Bandwidth Throttling:} With data traveling over the internet, there is a higher risk of interception, unauthorized access, and data leakage. By segmenting networks, using VPN tunnels, and limiting bandwidth, one reduces the chance of large-scale data theft. Even if a remote attacker gains foothold, the isolated and throttled nature of the network environment makes large-scale exfiltration difficult and more easily detectable.

    \item \greenbf{Immutable Logs and Blockchain:} Governance also involves proving compliance after the fact. Immutable logs stored on a blockchain provide an evidence trail of all activities. In a remote scenario, these records are vital for regulatory inquiries, ensuring that the entire evaluation - \textit{Who accessed what? Which computations were run? When results were generated?} - can be verified months or years later without relying on untrusted local logs. In the context of serious incident reporting (an obligation under the EU AI Act’s Code of Practice), the existence of such an evidence trail would be crucial in investigating incident causality\footnote{\url{https://www.pourdemain.ngo/en/post/learning-from-history-gpai-serious-incident-reporting}}, deploying appropriate corrective measures and factoring learnings into future risk mitigation.

    \item \greenbf{System-Wide Differential Privacy:} Even governance metrics, such as how often certain data was accessed or aggregated reports about system usage, can inadvertently reveal patterns. Applying differential privacy to these meta-level analytics ensures no sensitive patterns slip through the cracks. In remote evaluating systems that may span multiple jurisdictions, this helps align with privacy regulations while maintaining a broad overview of system health and compliance.
\end{itemize}

\subsection{Legal Safeguards}\label{legal}

As mentioned in Section \ref{onsite}, physical safeguards for secure external assessments, in the form of on-site evaluations, is the most comprehensive way of combining state-of-the-art deeper-than-black-box evaluations with the necessary security measures. Section \ref{cybersec}, practically details how such security measures could be ensured through technical means, before the necessary global regulatory frameworks are instituted to mandate on-site evaluations.

A third method, involving legal safeguards, can serve to bolster the security of external assessments, both on-site and remote. Such legal safeguards are an important complimentary fixture of established auditing practices, including in the financial sector, that provide an extra layer of protection for the containment of sensitive information. Such examples for external assessors include confidentiality and ‘handling of sensitive information’ training; a common practice in third-party auditing that may involve clear guidance on confidential and ‘non-confidential’ materials, training on securing digital and physical documents, training involving case studies of mishandling of information, breach simulations, and escalation pathways for auditors suspecting data breaches \cite{octaviani2021effect}. Other legal safeguards include contractual arrangements featuring non-disclosure clauses with clear penalties and timelines, standardized ‘terms of engagement’ between auditors and auditees (such as the scope of services, out-of-scope activities and mutual responsibilities \cite{asa2006auditing}), as well as provisions to avoid conflicts of interest \cite{casper2024black}.

Whilst such legal safeguards boost the security of the technical safeguards listed above, an important caveat is that this should not come at the expense of whistleblower protections.

\section{Recommendations, Further Action \& Conclusions}\label{part3}

The landscape of third-party external assessments for General-Purpose AI (GPAI) is rapidly evolving, necessitating robust governance frameworks that balance security, transparency, and technical feasibility. Our analysis underscores that black-box evaluation techniques are insufficient for rigorous AI audits, highlighting the \greenbf{need for deeper-than-black-box external assessments} with structured access to model internals. In this paper, we also detail the security vulnerabilities associated with remote external evaluations and propose a range of technical and legal safeguards to mitigate risks, ensuring that such audits remain viable in the absence of a globally recognized AI governance body.

\subsection{Policy Implications \& Governance Windows}

Several ongoing policy initiatives and governance structures provide avenues for implementing the recommendations outlined in this report. One such framework is the EU AI Act's \greenbf{Code of Practice for GPAI (CoP)}, which should explicitly endorse deeper-than-black-box access for external evaluations in its future iterations. The CoP can bridge the gap between voluntary best practices and eventual regulatory mandates by establishing structured frameworks for secure remote audits.

Another critical development is the rise of \greenbf{National AI Safety Institutes}, particularly in jurisdictions such as the UK and the US. These institutes are poised to play a pivotal role in certifying third-party auditors and formalizing security protocols for external assessments. Their work could help standardize auditing methodologies while ensuring compliance with national and international safety standards.

At a global level, \greenbf{international agreements} remain an essential yet fragmented component of AI governance. The emergence of AI-specific provisions in multilateral agreements—such as UN-led discussions and the Council of Europe Framework Convention—could lay the groundwork for harmonized auditing standards. While a single, unified regulatory body may not be imminent, coordinated policy efforts can drive meaningful progress toward establishing common principles.

Finally, \greenbf{industry-led initiatives} are an increasingly relevant mechanism for strengthening AI accountability. If AI labs are encouraged to adopt secure external auditing as a core part of their risk-management processes, companies can demonstrate a commitment to transparency while simultaneously advancing best practices in AI safety.

\subsection{Next Steps for Implementation}

To translate these findings into concrete action, several critical questions must be addressed. The first concerns \greenbf{what types of tests should be conducted} to enhance AI safety and reliability. Capability and safety probing will be essential to evaluate models for different dangerous propensities: from deception to power-seeking behavior, through long-horizon instrumental goals, we have outlined several state-of-the-art safety techniques that could be employed in a deeper-access model assessment.

The second key question revolves around \greenbf{who should conduct these evaluations}. A hybrid, multi-stakeholder approach is necessary, one that leverages the distinct strengths of both public and private assessors \cite{oueslati2025external,trager2025who}. Public evaluators—such as dedicated units within the AI Office—offer greater independence, the authority to secure sensitive model access, and are particularly well-suited for domains like national security where strict oversight is crucial. In contrast, private external assessors, similar to the EU’s Conformity Assessment Bodies (CABs), provide scalable expertise and specialized technical competence across various risk domains. Moreover, regulatory agencies and AI Safety Institutes must play an active role in setting audit criteria and certification requirements. To further ensure robustness and impartiality, academic institutions and civil society organizations should contribute by offering independent oversight and methodological diversity. In cases where contractual relationships exist between assessors and model providers, stringent rules and public oversight are essential to avoid conflicts of interest. Thus, aligning with others, we propose a hybrid model: \greenbf{combining public oversight with the technical agility of private bodies}; this model aims to deliver comprehensive evaluations that address the full spectrum of risks associated with GPAI models.

Equally important is the issue of access, as \greenbf{assessing different model propensities requires varying levels of access}. While some behaviors, such as overt biases or robustness under distribution shifts, can be evaluated with black-box access through API interactions, more complex risks—such as deception, power-seeking tendencies, or latent goal misgeneralization—necessitate deeper access. However, this does not mean that all external assessments must be white-box ones. As we have argued in this piece, \greenbf{evaluators should be granted the appropriate depth of access}, ranging from \textit{de facto} white-box access for scrutinizing internal representations to more restricted API-level testing when full access is not feasible but targeted safety evaluations remain necessary.

\subsection{Broader Impact: Safety \& Scientific Advancements}

Beyond ensuring rigorous safety guarantees, \greenbf{deeper-than-black-box auditing} methodologies have the potential to \greenbf{advance the broader science of AI interpretability and robustness}. Structured third-party access to AI models would facilitate early detection and mitigation of systemic risks, reducing the likelihood of catastrophic failures or misuse. Furthermore, increased transparency in AI evaluations would enhance public trust, reinforcing the credibility of GPAI deployments across sectors.

The adoption of these methodologies could also \greenbf{accelerate advancements in AI evaluation research}. Encouraging the development of novel techniques for structured auditing would refine existing methodologies, paving the way for more comprehensive assessments of AI behavior. This, in turn, would contribute to a safer, more accountable AI ecosystem—one where security, transparency, and responsibility are embedded as foundational principles from the outset.

\subsection*{Conclusion}

This paper outlines a \greenbf{scalable, secure, and transparent framework for external AI evaluations}, bridging the gap between current technical capabilities and emerging governance frameworks. While black-box assessments remain a valuable tool, the next frontier of AI safety requires embracing structured access models that balance security, transparency, and accountability. The immediate priority is to \greenbf{integrate these findings into existing policy windows}, ensuring that future iterations of the Code of Practice, AI Safety Institutes, and international agreements embed these principles into binding governance frameworks.

We suggest that relevant \greenbf{stakeholders develop structured access protocols}, secure auditing infrastructures, and accreditation mechanisms for third-party evaluators, so that we can lay the foundation for a more robust and accountable AI ecosystem—one where safety and trustworthiness are not afterthoughts, but core design principles from the outset.

\section*{Acknowledgements}
\addcontentsline{toc}{section}{Acknowledgements}

We would like to thank Jasmine Brazilek, Amin Oueslati, Toni Lorente, Afek Shamir and Julia Mykhailiuk for their detailed and useful feedback on early versions of this manuscript. We are also grateful to Stephen Casper and Ben Bucknall for their helpful suggestions.

\sloppy

\addcontentsline{toc}{section}{References}

\end{document}